# Crossover from 2D ferromagnetic insulator to wide bandgap quantum anomalous Hall insulator in ultra-thin $MnBi_2Te_4$

Chi Xuan Trang[1,2#], Qile Li[1,2,3#], Yuefeng Yin[2,3#], Jinwoong Hwang [4], Golrokh Akhgar[1,2], Iolanda Di Bernardo[1,2], Antonija Grubišić-Čabo[1], Anton Tadich[5], Michael S. Fuhrer[1,2], Sung-Kwan Mo[4], Nikhil Medhekar[2,3*], Mark T Edmonds[1,2*]

[1]School of Physics and Astronomy, Monash University, Clayton, VIC, Australia.

[2]ARC Centre for Future Low Energy Electronics Technologies, Monash University, Clayton, VIC, Australia.

[3]Department of Materials Science and Engineering, Monash University, Clayton, VIC, Australia

[4]Advanced Light Source, Lawrence Berkeley National Laboratory, Berkeley, CA, USA.

[5]Australian Synchrotron, 800 Blackburn Road, Clayton, VIC, Australia.

# These authors contributed equally
* Corresponding author: mark.edmonds@monash.edu and nikhil.medhekar@monash.edu

**Abstract**

**Intrinsic magnetic topological insulators offer low disorder and large magnetic bandgaps for robust magnetic topological phases operating at higher temperatures. By controlling the layer thickness, emergent phenomena such as the quantum anomalous Hall (QAH) effect and axion insulator phases have been realised. These observations occur at temperatures significantly lower than the Néel temperature of bulk $MnBi_2Te_4$, and measurement of the magnetic energy gap at the Dirac point in ultra-thin $MnBi_2Te_4$ has yet to be achieved. Critical to achieving the promise of this system is a direct measurement of the layer-dependent energy gap and verification of a temperature-dependent topological phase transition from large bandgap QAH insulator to a gapless TI paramagnetic phase. Here we utilise temperature dependent angle-resolved photoemission spectroscopy to study epitaxial ultra-thin $MnBi_2Te_4$. We directly observe a layer dependent crossover from a 2D ferromagnetic insulator with a bandgap greater than 780 meV in one septuple layer (1 SL) to a QAH insulator with a large energy gap (>70 meV) at 8 K in 3 and 5 SL $MnBi_2Te_4$. The QAH gap is confirmed to be magnetic in origin, as it becomes gapless with increasing temperature above 8 K.**

Topological insulators (TIs) in three-dimensions possess a topologically protected spin-polarized gapless Dirac cone on the surface of a bulk insulator that is robust against time-reversal invariant perturbations[1-7]. Time-reversal symmetry in TIs can be broken by introducing long range magnetic order, resulting in profound changes to the electronic band structure, specifically a gap opening at the Dirac point caused by exchange coupling[6,8-10]. The combination of magnetization and strong spin-orbit coupling in ultra-thin topological insulators not only results in a gap opening but also the presence of chiral edge modes within the magnetic gap representing the quantum anomalous Hall (QAH) insulating state[9-11]. In this state the edge mode is chiral and perfectly spin polarized, yielding dissipationless transport of charge with applications in spintronic and ultra-low energy electronics[12]. Further, realization of new topological phases based on broken time reversal symmetry such as the chiral Majorana fermion[13] and axion insulator states are also achievable[14,15].

The QAH effect was first realised via dilute magnetic doping using Cr and V in ultra-thin films of $(Bi,Sb)_2Te_3$ [9,10]. However, the non-uniform doping and magnetization result in large spatial fluctuations in the size of the magnetic gap[16], and the parallel dissipative channels create spin-scattering which greatly suppresses the temperature at which QAH effect can be observed. Modulation-doped sandwich heterostructures have modestly raised the QAHE temperature to 1-2 K[17,18], which is still far below the size of the magnetic gap and the Curie temperature. To explore the QAHE and other topological phases at elevated temperatures requires uniform distribution of magnetisation, not readily achieved via dilute magnetic doping.

The intrinsic magnetic topological insulator $MnBi_2Te_4$ was recently proposed[19-22] and experimentally verified as a bulk antiferromagnetic topological insulator[23-26] that hosts both intrinsic magnetism and topological protection. $MnBi_2Te_4$ is a layered compound similar to the well-known TI $Bi_2Te_3$, where five atomic layers of Te - Bi -Te - Bi - Te form a quintuple layer (QL), however, $MnBi_2Te_4$ has an extra Mn - Te layer in between Te and Bi in the middle of the $Bi_2Te_3$ quintuple layer, forming a septuple layer (SL) (Fig. 1a). The origin of the magnetic order comes from the $Mn^{2+}$ ions (contributing a 5 $\mu_B$ magnetic moment), with moments coupled ferromagnetically within each SL, and antiferromagnetically between adjacent SLs, resulting in an uncompensated antiferromagnet. By controlling the layer structure in the 2D limit a remarkable set of thickness-dependent magnetic and topological transitions have been predicted, such as wide bandgap ferromagnetic (FM) insulator in 1 SL with potential applications in proximity-induced magnetisation [21]. Beyond 1 SL, even- and odd-layered systems are predicted to be axion insulators and wide-bandgap QAH insulators respectively[19,22]. Recent transport experiments on $MnBi_2Te_4$ have confirmed the axion insulator phase in 6 SL [15] and the QAHE in 5 SL at 1.4 K at zero magnetic field, further increased to 6.5 K in an external magnetic field by aligning the layers ferromagnetically[27]. Yet these temperatures correspond to thermal energies <1 meV and are still well below the predicted magnetic bandgap values (ranging between 38 meV[19] and 98 meV[27]) and below the Néel temperature ($T_N \approx$ 23-25 K[25,27]). To date, the electronic bandstructure of ultra-thin $MnBi_2Te_4$ has only been examined above the Néel temperature where it appears gapless.[28] Hence there has been no direct confirmation of the crossover from ferromagnetic (FM) insulator to QAH insulator with increasing layer thickness, nor direct observation of the size or magnetic nature of the energy gap in the QAH state.

In this work we precisely control the layer thickness of $MnBi_2Te_4$ via molecular beam epitaxy growth (MBE) and perform temperature-dependent angle-resolved photoelectron spectroscopy (ARPES) above and below the Néel temperature to reveal the layer-dependent crossover from 2D ferromagnetic insulator ($\Delta_{GAP} >$ 780 meV) in 1 SL to QAH insulator with large gap ($\Delta >$ 70 meV at 8 K) in 3 and 5 SL, in excellent agreement with first principles density functional theory (DFT) calculations. The QAH gap is confirmed to be magnetic in origin, as it abruptly becomes zero with increasing temperature as the system becomes paramagnetic, demonstrating a phase transition from QAH insulating state to topological gapless surface state.

High-quality $MnBi_2Te_4$ films were grown by molecular beam epitaxy on a Si(111)-7 × 7 substrate via alternately growing 1 QL $Bi_2Te_3$ and 1 BL MnTe to spontaneously form $MnBi_2Te_4$ (see Section 1 of SI for specific details on the growth). Figure 1b shows a Reflection High-energy Electron Diffraction (RHEED) pattern of 3 SL $MnBi_2Te_4$. The atomically flat morphology of the film is indicated by a sharp (1 × 1) streak pattern (see SI Section 1 for LEED and additional RHEED data). Figure 1c shows an atomic-resolution scanning tunnelling microscopy (STM) image and the expected 1 × 1 atomic structure with lattice constant of 4.3 Å. A constant binding energy ARPES map taken at the Fermi level (i.e. Fermi surface map) of 2 SL $MnBi_2Te_4$, capturing multiple Brillouin zones (BZ) is shown in Fig. 1d. In each BZ, the only features observed are at the $\bar{\Gamma}$ points, showing a slight hexagonal warping of the Fermi surface. The hexagonal features confirm threefold rotational symmetry, consistent with the rhombohedral structure of $MnBi_2Te_4$. An in-plane lattice constant, $a \approx 4.3$ Å can be calculated from the Fermi surface map for 2 SL, with similar values for 3 and 5 SL (see SI Section 2) in excellent agreement with STM (Fig. 1c) and previous reports on bulk $MnBi_2Te_4$[26,29]. In contrast, 1 SL has a lattice constant of 4.06 Å suggesting there is compressive strain when in direct contact with the Si(111), immediately relaxed when the thickness increases. These results are summarized in Table 1. Figure 1e shows the angle-integrated photoemission spectrum of 5 SL taken at $h\nu$ = 100 eV, with the characteristic Mn 3*p*, Te 4*d* and Bi 5*d* core levels observed (detailed XPS measurements are shown in Fig. S3).

By delicate control of the growth conditions, we achieve SL-by-SL growth of $MnBi_2Te_4$, allowing the DFT-predicted transition from ferromagnetic insulator to quantum anomalous Hall insulator to be probed directly with ARPES. The upper panel of Figure 2 shows a series of ARPES band maps of $MnBi_2Te_4$ with thicknesses of (a) 1 SL, (b) 2 SL, (c) 3 SL and (d) 5 SL measured along the $\bar{\Gamma}\bar{M}$ direction. The lower panel in Fig. 2 plots the respective double derivative, overlaid with the corresponding DFT calculations using HSE functional (1-3 SL) and PBE functional (5 SL). There is a clear thickness dependent transition from wide bandgap in 1 SL (Fig. 2a) to an evolution into nearly Dirac-like dispersion in 3 and 5 SL (Fig. 2c-d), evidence of an evolution towards non-trivial topological features. There is an excellent overlap with the DFT calculations in the overall band shape and with increasing thickness, along with a rigid shift in the bands toward the Fermi level, suggesting the film becomes less *n*-type due to increased manganese content which acts as a hole dopant.[6]

In Fig. 3a-b we determine the bandgap in 1 and 2 SL $MnBi_2Te_4$ respectively, by plotting the ARPES spectrum (left panel), its double derivative (middle panel) and the corresponding energy distribution curves (right panel). In Fig. 3a the ARPES spectrum of 1 SL $MnBi_2Te_4$ exhibits only a broad M-shaped valence band (VB), with the valence band maximum ≈780 meV below the Fermi level, with no signature of the conduction band (CB) and a strong intensity bulk Si(111) band observed below 1.1 eV. The overall band shape confirms an indirect bandgap ferromagnetic insulator in excellent agreement with our DFT results and Ref. 22. This is different to that reported by Ref. 19 which reports a direct bandgap, which is likely as a result of applying the on-site Hubbard correction term to Te instead of Mn (further discussion and calculations are found in Section S4 of the SI). The gap >780 meV is larger than theoretical predictions (see Table 1) and is either due to bandgap underestimation in DFT[30] or the compressive strain that is observed in 1SL. The gap size is comparable to other 2D ferromagnets, such as monolayer $CrI_3$[31], but is significantly larger than in (non-magnetic) 1 QL pristine $Bi_2Te_3$[4]. In Fig. 3b for the 2 SL film the M-shaped VB is preserved but now a parabolic

CB appears, whose minimum is below $E_F$, indicating the film is *n*-type doped. The CB dispersion within a 0.15 Å$^{-1}$ region extending from the $\bar{\Gamma}$ point is fit using a nearly free electron model, $E = \hbar^2 k^2 / 2m^*$, yielding an effective mass, $m^* = 0.25\ m_0$ (see Section S5 and Fig. S7 in SI for fits). An extremely weak intensity hole band, only observable when taking a double derivative (middle panel of Fig. 3b), results in a direct bandgap of 300 ± 100 meV. The excellent agreement with DFT suggests that 2 SL is a zero plateau QAH insulator.[22]

We now examine the size of the bandgap in 3 SL and 5 SL $MnBi_2Te_4$. Figure 4a shows high-resolution ARPES maps taken at $h\nu$ = 50 eV and 8 K (i.e. well below the Néel temperature, $T_N$ ≈ 23-25 K[25,27]) for 3 SL (top panel) and 5 SL (bottom panel). The red circles overlaid reflect the extracted maxima from fitting momentum distribution curves (MDC) with Lorentzian line shapes in the $k$ range between -0.5 to +0.5 $Å^{-1}$ as shown in Fig. 4b. Fitting with four line shapes was used for binding energies below ~0.4 eV and ~0.3 eV for 3 SL and 5 SL respectively, in order to decouple the linearly dispersing Dirac bands from additional valence bands which are clearly present in the MDCs in Fig. 4b and predicted in the DFT calculations in Fig. 4c. These additional bands when interacting with the Dirac hole band lead to an anti-crossing and manifest as a slight kink in the band dispersion. Strong spectral weight near $\bar{\Gamma}$ in the Dirac point region is due to Te-orbital-related matrix elements effects (see Fig. S6 for orbital character analysis) as this intensity diminishes quickly with changing photon energy (see Fig. S8). This strong spectral weight has previously been observed in the Dirac point region of $Bi_2Te_3$[3] and $Bi_2Te_3/MnBi_2Te_4$ heterostructures[7]. A clear electron and hole band asymmetry is observed, yielding asymptotic Fermi velocities for 3 SL of $v_{F,e} \approx 4.2 \times 10^5$ ms$^{-1}$ and $v_{F,h} \approx 2.0 \times 10^5$ ms$^{-1}$ and for 5 SL $v_{F,e} \approx 5.0 \times 10^5$ ms$^{-1}$ and $v_{F,h} \approx 2.9 \times 10^5$ ms$^{-1}$. The hole band $v_F$ for both 3 SL and 5 SL was extracted above the kink regions discussed above.

The band dispersion for both 3 SL and 5 SL appear hyperbolic as expected for a massive Dirac dispersion, and the linear extrapolation of the electron and hole bands to $k = 0$ shows the bands do not meet at a discreet point (i.e. the Dirac point) but are actually separated by 40 meV for both 3 SL and 5 SL $MnBi_2Te_4$. This strongly supports the conclusion that the system is gapped, and in order to accurately determine the size of the bandgap we utilize two independent methods. The first is based on the hyperbolic band dispersion for both 3 SL and 5 SL as expected for a massive Dirac dispersion, so we fit the data to a model of a hyperbolic system (shown as white bands in Fig. 4a) given by

$$E_i(k) = D \pm \sqrt{\Delta_i^2 + \hbar^2 v_{F,i}^2 (k + k_0)^2}, i \in n, p \quad (1)$$

where $\Delta = \Delta_n + \Delta_p$ represents the bandgap, D the doping, and $v_{F,i}$ the asymptotic Fermi velocities away from the gapped region at large momenta. This yields magnetic gap sizes of $\Delta_{3SL}$ = 109 ± 15 meV and $\Delta_{5SL}$ = 84 ± 15 meV. Figure 4c overlays the DFT band calculations and shows excellent overlap for 3 SL, whilst the 5 SL PBE (due to underestimation of the bandgap[30]) required the bandgap to be adjusted to match the experimental data using the scissors method.

A check on the determination of the bandgap was performed by peak fitting analysis of the energy distribution curves (EDCs) taken at $k_{||}$ = 0. The EDCs for 3 SL and 5 SL $MnBi_2Te_4$ taken at 8 K are plotted in Fig. 4d, along with the peak fitting analysis based on Ref. 7 for $Bi_2Te_3/MnBi_2Te_4$ heterostructures where a pronounced shoulder is observed when cooling below 13 K (the temperature dependence and the evolution of electronic structure is shown in Fig. 5 and discussed in detail below). We fit the dominant spectral weight with two identical Lorentzian line shapes, and include small additional conduction and valence band peaks either side of the main peaks that reflect the electron and hole Dirac bands. At 8 K we arrive at gaps for 3 SL and 5 SL of 71 ± 15 meV and 70 ± 15 meV respectively. The bandgap derived from EDC fitting shows excellent agreement with that obtained from fitting to Eqn. (1) for 5 SL $MnBi_2Te_4$ and slight discrepancy above experimental error for 3 SL. Figure 4e plots the bandgap as a function of thickness for experiment and DFT calculations from this work and that of Ref. 22, and are also summarised in Table 1.

We now investigate the response of the topological electronic structure to magnetic ordering, by conducting temperature-dependent ARPES measurements. Figure 5a shows the energy distribution curves at $k_{||}$ = 0 measured at temperature $T$ = 8 K and 13 K for 5 SL $MnBi_2Te_4$. There is a clear broadening upon cooling from 13 K, and the right flank of the peak (i.e. lower binding energy) develops a clear shoulder. Higher temperature measurements (up to 33 K) show minimal change from the 13 K data, which suggests that 8 K is below the magnetic phase transition temperature, i.e. uncompensated antiferromagnetic phase (uAFM), and 13 K above it, i.e. paramagnetic phase. In Fig. 5b we fit the $T$ = 13 K spectra (fitting up to 33 K was also performed) with two almost identical components yielding a splitting of 15 ± 15 meV. A slightly larger peak intensity for the hole band was adopted due to the lower Fermi velocity of the hole band. This splitting of 15 ± 15 meV reflects a nearly gapless topological insulator (that maintains a small remnant confinement gap due to its ultra-thin nature) confirming the system is now paramagnetic. The sum of these two peaks is plotted as a single peak component (red line shape) for clarity in Fig. 5b. Upon cooling to 8 K and the uAFM phase, the peak splitting of these two components (blue line shapes) increases and yields a magnetic gap, Δ= 70 ± 15 meV.

There is further evidence of a band evolution between 8 and 13 K when examining the near-$E_F$ ARPES maps and the corresponding MDCs at 8 K (Fig. 5c) and 13 K (Fig. 5d) (full temperature range 8-33 K is shown in Fig. S9). As shown in Fig. 5e at binding energies, $E_B$ = 0, 0.25, and 0.36 eV there is a clear increase in wavevector, $k$ with increasing $T$. However, the slope of the electron and hole band (and consequently Fermi velocity) extracted from the MDC maxima remains unchanged as shown in Fig. S9c of the SI. At 13 K and above linear extrapolation of the bands to $k = 0$ yields a gap $\Delta < 10$ meV in excellent agreement with the EDC analysis above. The near gapless nature at 13 K and above is further confirmed via ARPES measurements taken at $T$ = 13 K and $hv$ = 40 eV (shown in Fig. S10) in order to avoid the strong spectral weight of the Te orbital in the gapped region. This clearly shows a gapless Dirac dispersion and is consistent with measurements taken on 5 and 7 SL $MnBi_2Te_4$ at 25 K in Ref. 28. Figure 5f plots the bandgap as a function of temperature.

This clear emergence of a magnetization induced gap with decreasing temperature provides a definitive smoking-gun signature for a temperature-dependent topological phase transition from large bandgap QAH insulator to a near gapless TI paramagnetic phase. These phases are depicted schematically in Fig. 5g. It should be noted the phase transition temperature reported here between 8-13 K is well below the $T_N \approx 23$ K reported for 5 SL $MnBi_2Te_4$ [27] however, in both Ref. 22 and 27 a decrease in the Néel temperature is reported with decreasing thickness. Furthermore, Ref. 27 reports corrections to the temperature dependent resistivity with an abrupt downturn in resistivity below $T_N$, followed by a rapid increase at ≈10 K which they attribute to localization. The abrupt transition observed at similar temperature here in ARPES, along with the QAH effect in $MnBi_2Te_4$ being limited to 6.5 K [27], suggests that the phase transition responsible for topological order may occur below the Néel temperature (indicating more than one magnetic phase transition) or the Néel temperature may be lower in these samples than previously thought and further work is needed to understand this behaviour.

Our results provide the first experimental demonstration of the unique thickness-dependent electronic properties of $MnBi_2Te_4$, from a wide bandgap 2D ferromagnet to a QAH insulator with a magnetic bandgap in excess of 70 meV. Therefore, $MnBi_2Te_4$ not only offers pathways to realise high temperature QAH effect but is also applicable in designer van der Waals heterostructures for proximity induced magnetization[21] and in the realization of new topological phases such as chiral Majorana fermions via coupling to a superconductor[13].

**Acknowledgements**

M.T.E. was supported by ARC DECRA fellowship DE160101157. M.T.E., C.X.T., Q.L., Y. Y., I. B., G. A., N. M., M.S.F. acknowledge funding support from CE170100039. M.T.E. and C.X.T. acknowledge travel funding provided by the International Synchrotron Access Program (ISAP) managed by the Australian Synchrotron, part of ANSTO, and funded by the Australian Government. M.T.E., C.X.T., and Q.L. acknowledge funding support from ARC Centre for Future Low Energy Electronics Technologies (FLEET). Y.Y. and N.V.M. are thankful for the computational support provided by the Monash Computing Cluster, the National Computing Infrastructure and the Pawsey Supercomputing Facility. This research used resources of the Advanced Light Source, which is a DOE Office of Science User Facility under contract no. DE-AC02-05CH11231. Part of this research was also undertaken on the Soft X-ray beamline at the Australian Synchrotron, part of ANSTO.

**Author contributions** M.T.E. devised the experiments. Q. L., C. X. T., G. A performed the MBE growth at Monash University. C. X. T., Q. L., M.T.E., A. G. C., and I. B. performed the MBE growth and ARPES measurements at the ALS with the support from J.H. and S.-K.M. I. B. performed the STM measurements at Monash University. The DFT calculations were performed by Y. Y., and N. M. C. X. T., Q. L., and M.T.E. composed the manuscript. All authors read and contributed feedback to the manuscript.

**Competing interests** The authors declare no competing interests.

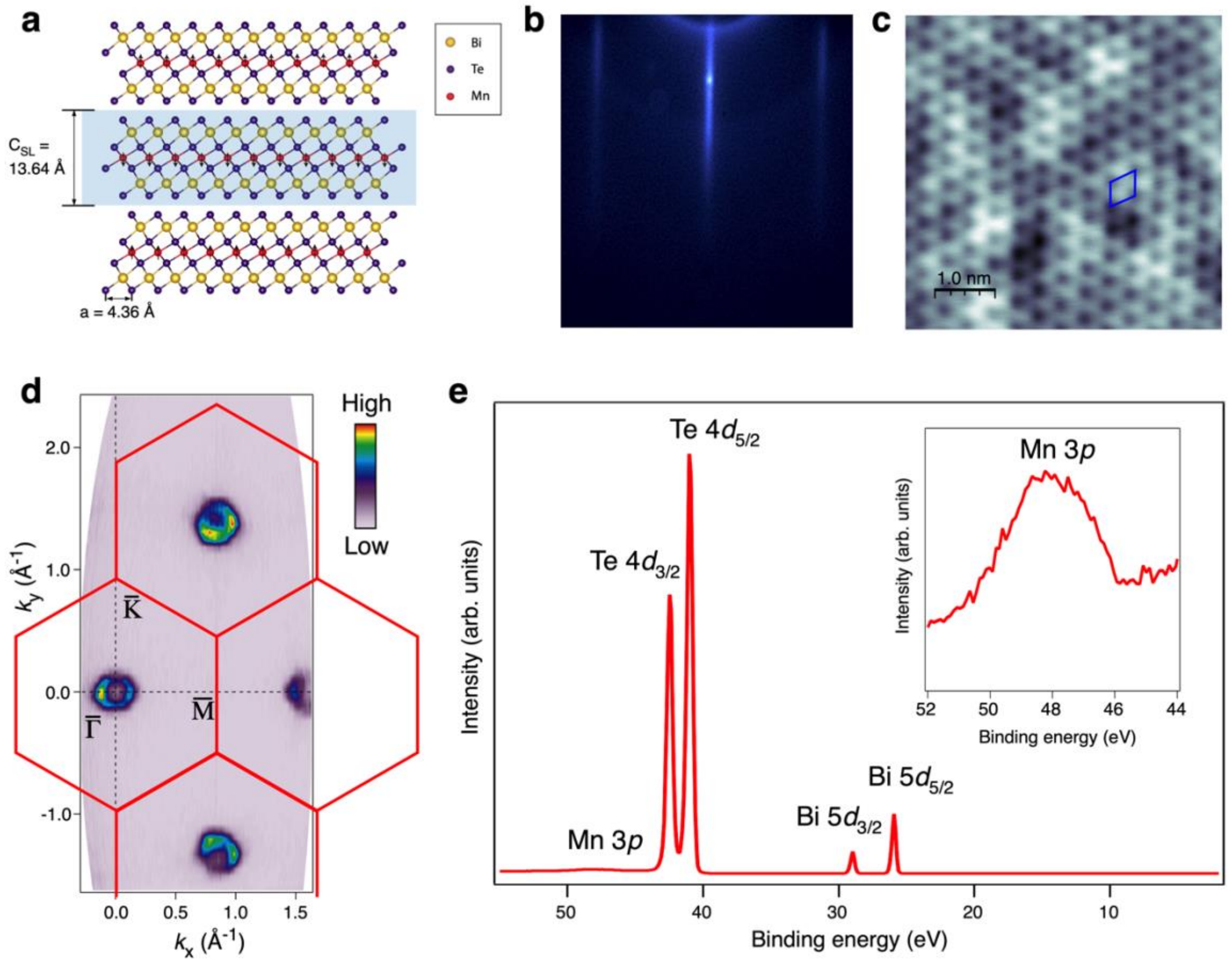

**Figure 1 | Basic properties and characterization of $MnBi_2Te_4$. a,** layered crystal structure of $MnBi_2Te_4$. **b**, Reflection High-Energy Electron Diffraction (RHEED) of 3 SL $MnBi_2Te_4$ along [112] direction. **c**, Atomic resolution scanning tunnelling microscope (STM) (5nm × 5nm) image of $MnBi_2Te_4$ (V = -2.7 V, I = 70 pA) showing the 1 × 1 structure (blue diamond). Lattice constant is 0.43 nm. **d**, Constant energy ARPES contour of 2 SL $MnBi_2Te_4$ taken at the Fermi level over multiple Brillouin zones (BZ). Hexagonal BZ is overlaid in red. **e,** Core-level spectrum of $MnBi_2Te_4$ taken at $h\nu$ = 100 eV showing the characteristic Mn 3*p*, Te 4*d* and Bi 5*d* core levels.

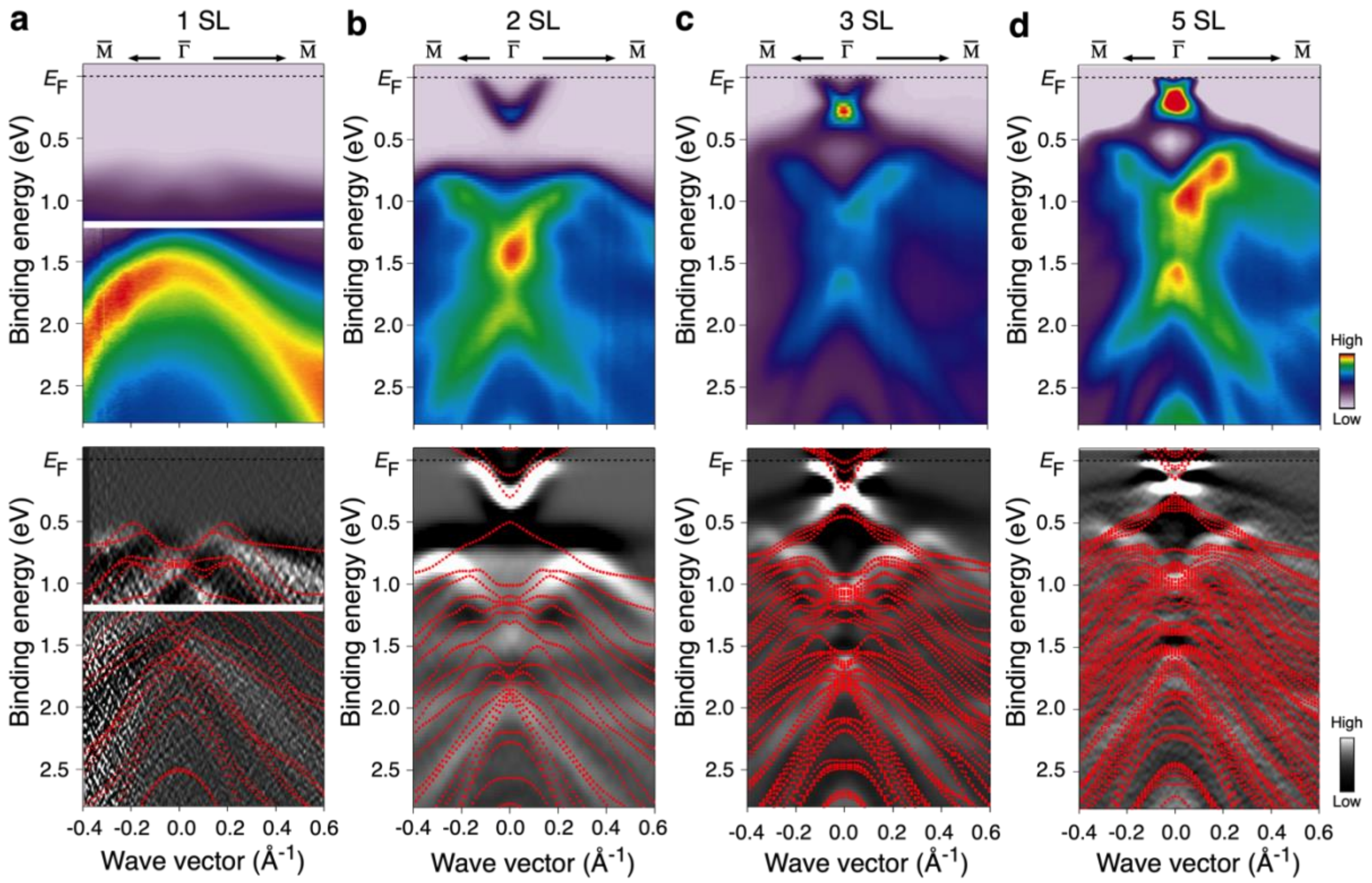

**Figure 2 | Thickness dependence of the band structure of $MnBi_2Te_4$ and comparison with calculation**. ARPES intensity (top) and double derivatives with overlaid DFT calculation (red dashed lines) (bottom) of **a**, 1 SL; **b**, 2 SL; **c**, 3 SL; and **d**, 5 SL along $\bar{\Gamma}\bar{M}$ cut, measured at $h\nu$ = 50 eV and $T$ = 13 K .

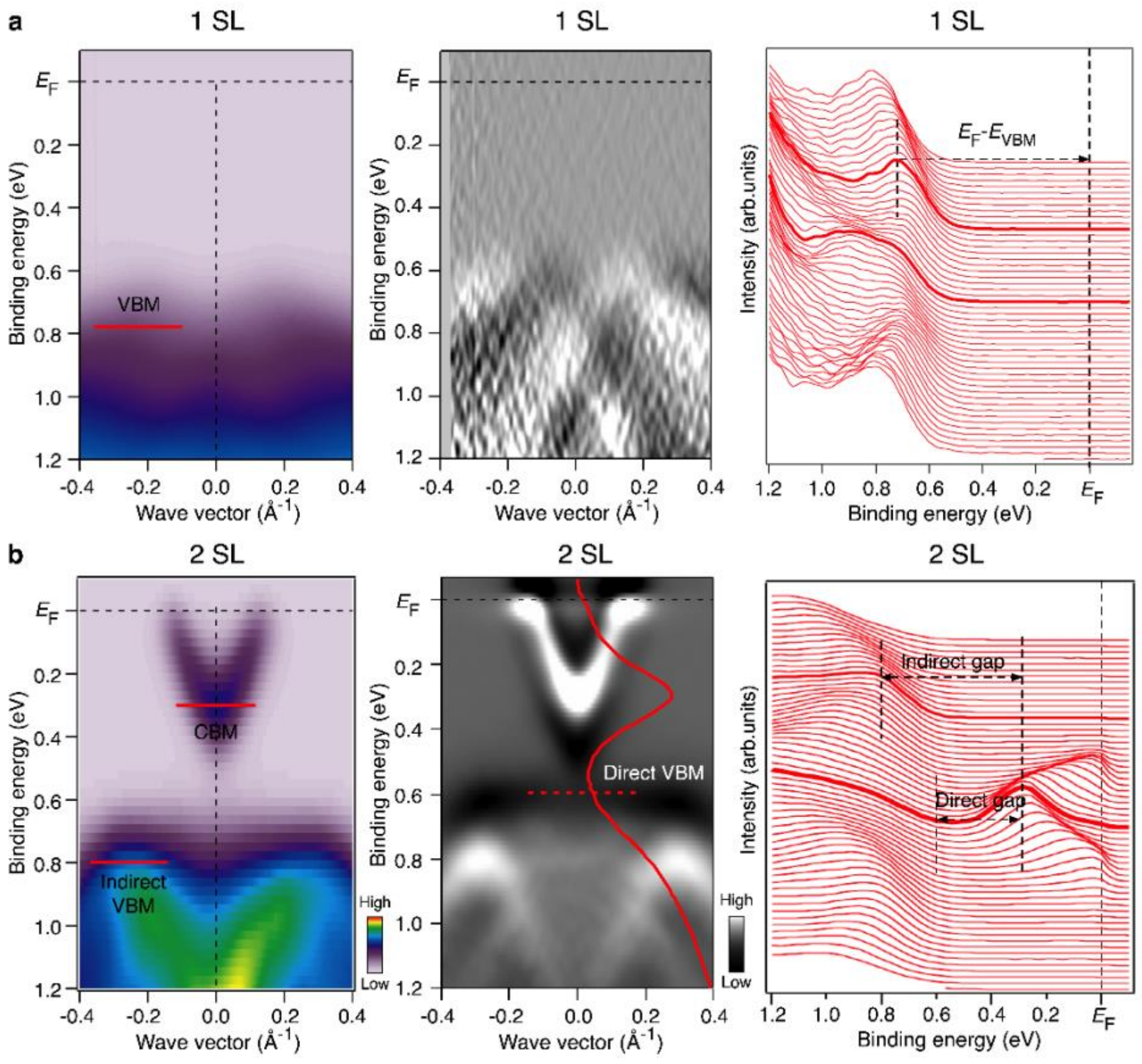

**Figure 3 | Bandgap of 1 and 2 SL $MnBi_2Te_4$.** Near $E_F$-ARPES intensity (left) at $h\nu$ = 50 eV and $T$ = 12 K, double derivatives (middle), and corresponding energy distribution curves (EDCs) (right) around the $\bar{\Gamma}$ point for **a**, 1 SL, and **b**, 2 SL. The VB maximum and CB minimum are marked as red lines in ARPES intensity.

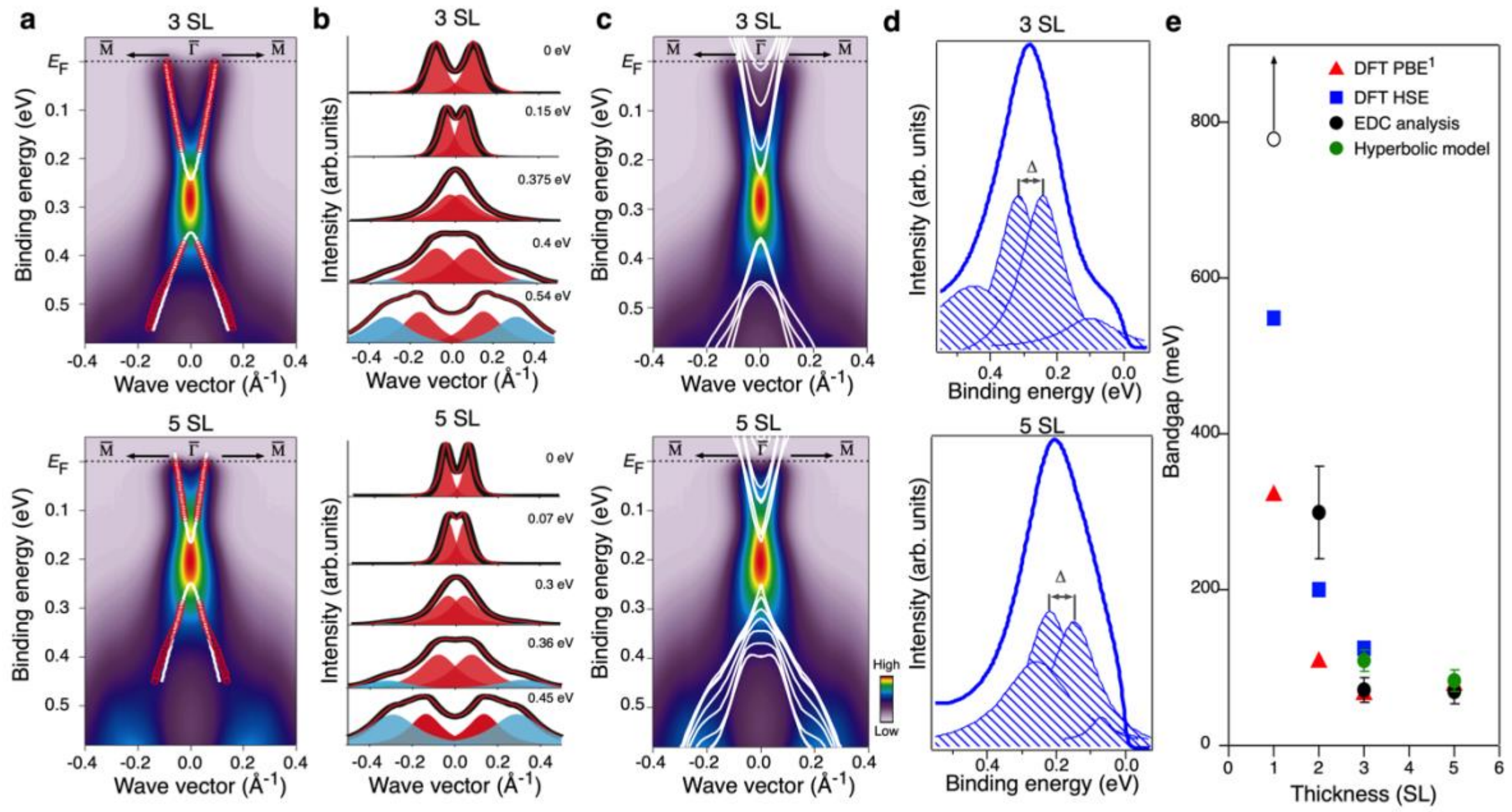

**Figure 4 | Magnetic gap of 3 and 5 SL $MnBi_2Te_4$.** High resolution ARPES intensity of 3 SL (top) and 5 SL (bottom) taken at $h\nu$ = 50 eV and $T$ = 8 K with overlaid **a**, hyperbolic band fittings (white lines) with extracted momentum distribution curve (MDC) peak maxima (red circles). **b**, MDCs at different $E_B$ fitted with Lorentzian line shapes. **c**, DFT calculations overlaid on the ARPES data. **d**, EDC spectra taken at $k_{||}$=0. The splitting of the two main peak line shapes corresponds to the magnetic gap, Δ. **e**, Bandgap as a function of thickness including data from experiment (black), fittings from Hyperbolic Model (green), DFT calculation (blue), and M. M. Otrokov *et al.*[22] (red). The open circle for 1 SL reflects only the $E_F$-$E_{VBM}$, and not the bandgap as the conduction band is not observed below the Fermi level.

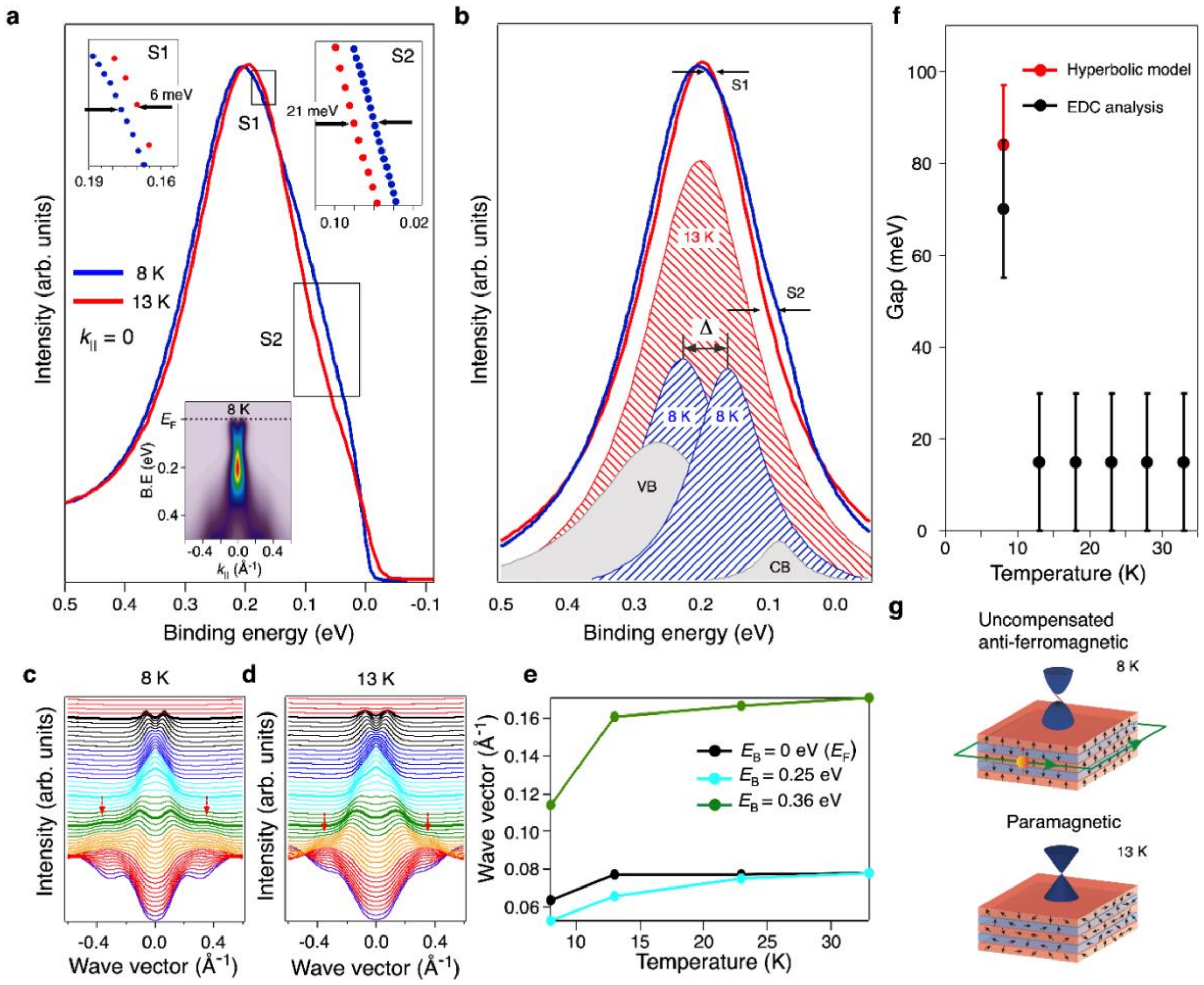

**Figure 5| Temperature dependence of 5 SL film**. **a**, Energy distribution curves (EDCs) taken at $k_{||}$=0 taken at 8 and 13 K for 5 SL $MnBi_2Te_4$. The regions S1 and S2 indicate a clear broadening and pronounced shoulder at 8 K, corresponding to a splitting of 6 meV and 27 meV, respectively. The inset in **a** shows the ARPES map taken at 8 K. **b**, Simulated peak fitting results from the spectra in **a**, corresponds to a magnetic gap of 70 ± 15 meV at 8 K and 15 ± 15 meV at 13 K. **c-d**, MDCs with different colours designating different binding energy regions for 3 SL and 5 SL respectively. Black: $E_F$-0.09 eV, blue: 0.1-1.9 eV, sky blue: 0.2-0.29 eV, green: 0.3-0.39 eV, orange: 0.4-0.49 eV, red: 0.5-0.59 eV, purple: 0.6-0.62 eV. **e**, Wave vector, *k* extracted from MDC peak maxima as a function of temperature at $E_B$ = 0 (black), 0.25 (blue), 0.36 (green) eV showing a clear band evolution with increasing temperature. **f**, Bandgap as a function of temperature. **g**, Schematics of uncompensated anti-ferromagnetism QAH insulator phase (upper) and paramagnetic gapless TI phase (lower).

**Table 1 Lattice constant, bandgap and Fermi velocity as a function of thickness of $MnBi_2Te_4$**

| | **Lattice constant, a (Å)** | **$E_g$ (meV) Theory** | **$E_g$ (meV) Experiment** | **$v_F$ (m/s) Hole** | **$v_F$ (m/s) Electron** |
|---|---|---|---|---|---|
| **1 SL** | 4.06 ± 0.30 | 550 [this work]<br>700 [Ref [19]]<br>321 [Ref [22]] | >780 ± 100 | - | - |
| **2 SL** | 4.31 ± 0.30 | 200 [this work]<br>107 [Ref [22]] | 300 ± 100 | - | - |
| **3 SL** | 4.31 ± 0.30 | 125 [this work]<br>66 [Ref [22]] | 109 ±15 (Hyperbolic Model)<br>71 ± 15 (EDC Model) | $2.0 \pm 0.5 \times 10^5$ | $4.2 \pm 0.5 \times 10^5$ |
| **5 SL** | 4.31 ± 0.30 | 38 [Ref [19]]<br>77 [Ref [22]]<br>98 [Ref [27]] | 84 ± 15 (Hyperbolic Model)<br>70 ± 15 (EDC Model) | $2.9 \pm 0.5 \times 10^5$ | $5.0 \pm 0.5 \times 10^5$ |

# Supplementary Information

## Crossover from 2D ferromagnetic insulator to wide bandgap quantum anomalous Hall insulator in ultra-thin $MnBi_2Te_4$

Chi Xuan Trang[1,2#], Qile Li[1,2,3#], Yuefeng Yin[2,3#], Jinwoong Hwang [4], Golrokh Akhgar[1,2], Iolanda Di Bernardo[1,2], Antonija Grubišić-Čabo[1], Anton Tadich[5], Michael S. Fuhrer[1,2], Sung-Kwan Mo[4], Nikhil Medhekar[2,3*], Mark T Edmonds[1,2*]

[1]School of Physics and Astronomy, Monash University, Clayton, Victoria, Australia.

[2]ARC Centre for Future Low Energy Electronics Technologies, Monash University, Clayton, Victoria, Australia.

[3]Department of Materials Science and Engineering, Monash University, Clayton, Victoria, Australia

[4]Advanced Light Source, Lawrence Berkeley National Laboratory, Berkeley, CA, USA.

[5]Australian Synchrotron, 800 Blackburn Road, Clayton, Victoria, Australia.

# These authors contributed equally
* Corresponding author: mark.edmonds@monash.edu and nikhil.medhekar@monash.edu

**Table of Contents**

1. **Experimental/theory methods and sample characterization with LEED and RHEED**

### Growth of Ultra-Thin $MnBi_2Te_4$ on Si(111)

Ultra-thin $MnBi_2Te_4$ thin films were grown in an ultra-high vacuum (UHV) molecular beam epitaxy (MBE) chamber and then immediately transferred after the growth to an interconnected ARPES chamber at Beamline 10.0.1, Advanced Light Source (ALS), Lawrence Berkeley National Laboratory, USA. To prepare an atomically flat substrate, a Si(111) wafter was flash annealed with e-beam heating in order to achieve a (7 × 7) surface reconstruction.

For $MnBi_2Te_4$ film growth, effusion cells were used to evaporate elemental Bi (99.999%) and Mn (99.9%) in an overflux of Te (99.95%). Rates were calibrated with a quartz crystal microbalance. High quality epitaxial growth of $MnBi_2Te_4$ was achieved by first growing 1QL of $Bi_2Te_3$ at 230°C, then a bilayer of MnTe was deposited in order to spontaneously form $MnBi_2Te_4$ in a similar manner to the formation of $MnBi_2Se_4$[1]. To reach the desired thickness this recipe was continued by again depositing 1 QL of $Bi_2Te_3$ followed by a bilayer of MnTe. After growth was completed the film was annealed at the same temperature in an overflux of Te for 5 - 10 min to improve crystallinity. Reflection High Energy Electron Diffraction (RHEED) and Low Energy Electron Diffraction (LEED) were used to confirm the (001) single-crystal epitaxial growth across a large area (see Supplementary Materials S1).

### Angle resolved photoemission spectroscopy (ARPES) measurements

ARPES measurements were performed at Beamline 10.0.1 of the ALS. Data was taken using a Scienta R4000 analyser at temperatures between 8 K and 33 K. The total energy resolution was 15 - 25 meV depending on the beamline slit widths and analyser settings, and the angular resolution was 0.2°. This resulted in an overall momentum resolution of ≈0.01 $Å^{-1}$ for the photoelectron kinetic energies measured, with the majority of the measurements performed at $h\nu = 50$ eV.

### Density Functional Theory Calculations

We employed density functional theory (DFT) calculations as implemented in the Vienna ab initio Simulation Package (VASP) to calculate the electronic structure of two-dimensional $MnBi_2Te_4$[2]. The electron exchange and correlation effects were treated with the Perdew-Burke-Ernzehof (PBE) form of the generalized gradient approximation (GGA)[3]. The kinetic energy cut-off for the plane-wave basis was set to 400 eV. We use a 12 × 12 × 1 Γ-centered *k*-point mesh for sampling the Brillouin zone. The van der Waals interactions in the system is described using the DFT-D3 potential[4]. To treat the strong, onsite Coulombic interactions of localized 3*d* electrons of Mn, which is inaccurately described by GGA, we used GGA+U approach with the effective Hubbard-like term U set to 4 eV[5]. The electronic bandstructure of 2D $MnBi_2Te_4$ obtained from PBE-GGA was further verified using the hybrid Heyd-Scuseria-Ernzerhof (HSE) hybrid functional[6].

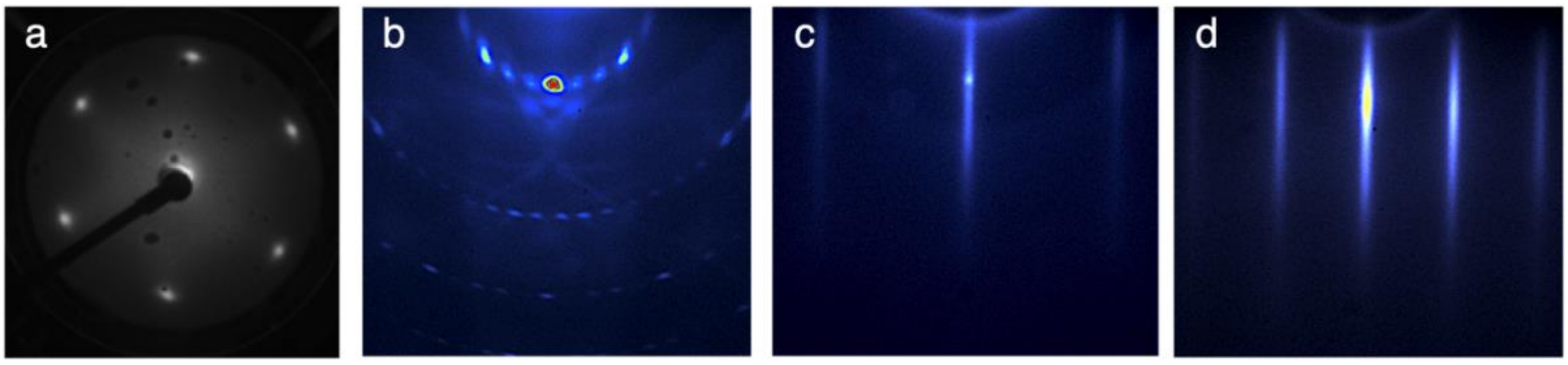

**Figure S3 | a**, Low-Energy Electron Diffraction (LEED) taken at 24 eV showing the 1 × 1 structure of $MnBi_2Te_4$. High-Energy Electron Diffraction (RHEED) showing **b,** Si(111)-7 × 7, and $MnBi_2Te_4$ along **c**, [112] and **d**, [110] directions.

## 2. Lattice constant as a function of thickness

The lattice constants are calculated from the formula $a = 4\pi/3d_{\Gamma K}$ or $a = 2\pi/\sqrt{3}d_{\Gamma M}$ where $d_{\Gamma K}$ or $d_{\Gamma M}$ is the distance between $\bar{\Gamma}$ and $\bar{K}/\bar{M}$, determined from the multiple Brillouin zones mapping as shown in Fig. S2. The thickness-dependent lattice constants are in Table 1 in the main text.

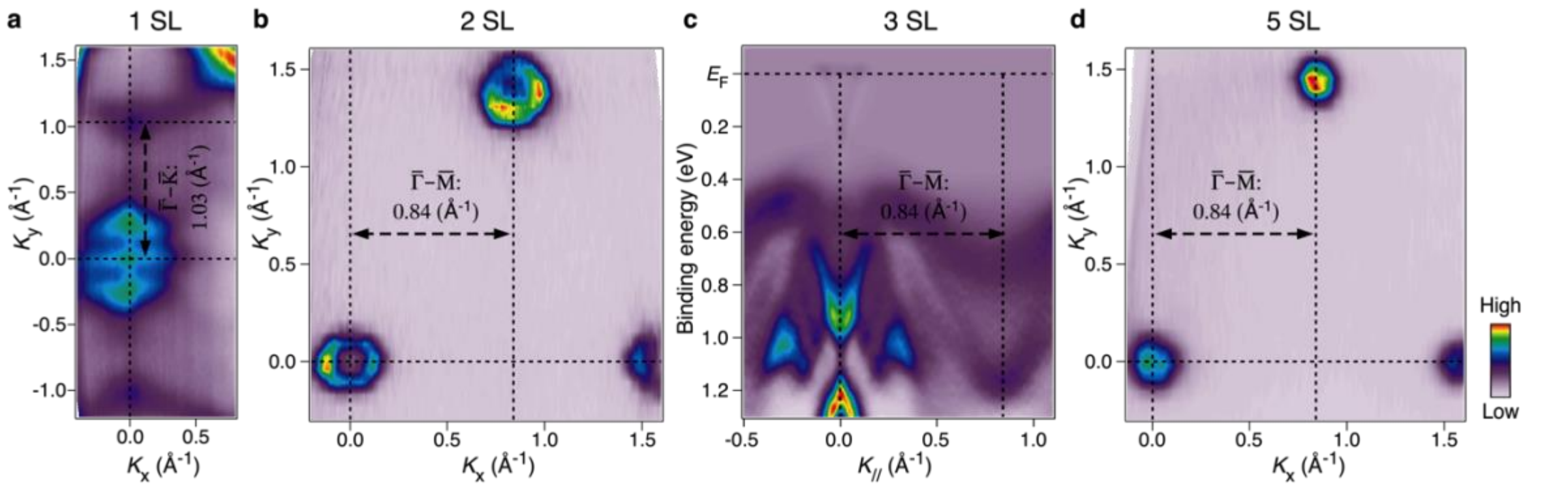

**Figure S4** | ARPES intensity mapping of **a**, 1 SL; **b**, 2 SL; **c**, 3 SL (along $\bar{\Gamma}\ \bar{M}$ cut); **d**, 5 SL.

## 3. X-ray photoelectron spectroscopy of $MnBi_2Te_4$

In order prove that the films are indeed $MnBi_2Te_4$, and to rule out a composition of $Bi_2Te_3$+MnTe, we used x-ray photoelectron spectroscopy to directly measure the binding energy of the Tellurium core level which is significantly different when Tellurium is bonded to Bismuth in $Bi_2Te_3$, compared to Tellurium bonded to Manganese in MnTe. In the literature, Kim *et al.*[7] report the growth of MnTe on Si(111) and measure XPS of the Te $3d_{5/2}$ core level at a binding energy of 571.4 eV. Our XPS measurements on $MnBi_2Te_4$ grown on Si(111) shown in the Fig. S3a below, reveal the Te $3d_{5/2}$ core level at 572.2 eV which is 0.8 eV higher in binding energy than MnTe. This separation is far larger than would be expected due to doping, and must be due to different bonding environments. Furthermore, if our films consisted of $Bi_2Te_3$+MnTe we would observe multiple peak components in XPS due to the different bonding environments. Figure S3b shows XPS at photon energies from 100 eV up to 1486 eV in order to increase the photoelectron mean free path and therefore the sampling depth of the experiment. At all photon energies we observe only single components in the Bi 5*d*, Te 4*d* and Mn 3*p* spin-split core levels, consistent with a bonding environment arising from $MnBi_2Te_4$ and not $Bi_2Te_3$+MnTe. These XPS measurements rigorously prove that our films are $MnBi_2Te_4$ and not $Bi_2Te_3$+MnTe.

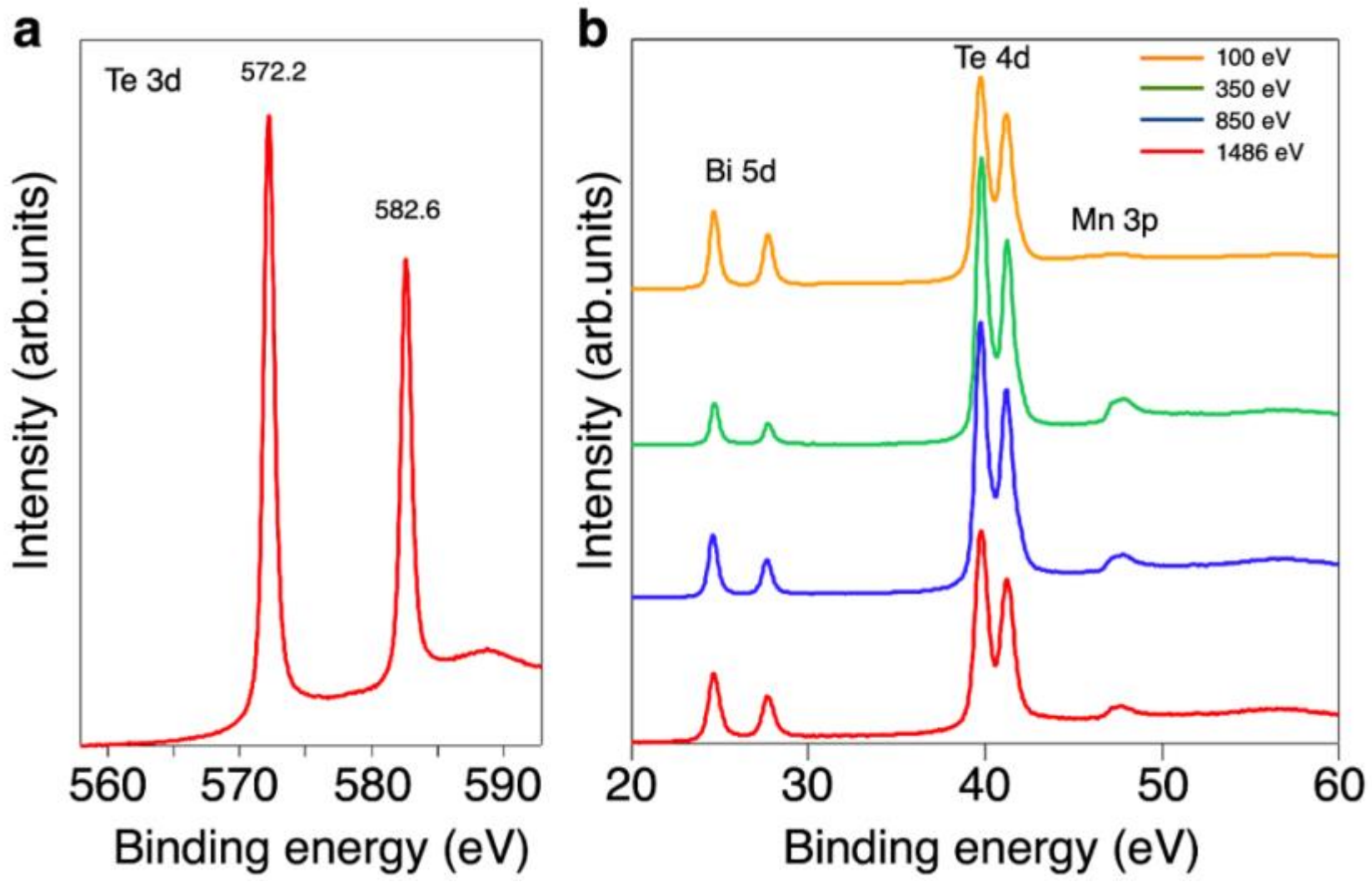

**Figure S3 |** X-ray photoelectron spectroscopy on $MnBi_2Te_4$ thin films. **a**, XPS of Te 3*d* core level. **b**, Depth dependent XPS of the Bi 5*d*, Te 4*d* and Mn 3*p* core levels taken at 100 eV (orange), 350 eV (green), 850 eV (blue) and 1486 eV (red).

## 4. (a) DFT calculations: Comparison between applying on-site Hubbard U correction term on Mn and Te

We have noticed that there are contrasting results in the literature on the band structure of $MnBi_2Te_4$ in previous DFT calculations of 1 SL $MnBi_2Te_4$ [8-10]. The differences can be attributed to the effect of applying on-site Hubbard U terms[5] for different atomic species (in this case, Mn and Te). Figure S4 shows a comparison between the band structures when applying U for Mn vs. Te. We find that U for Mn (Figure S3a) provides a better agreement with experimental observations, accurately reflecting the M-shaped valence band top.

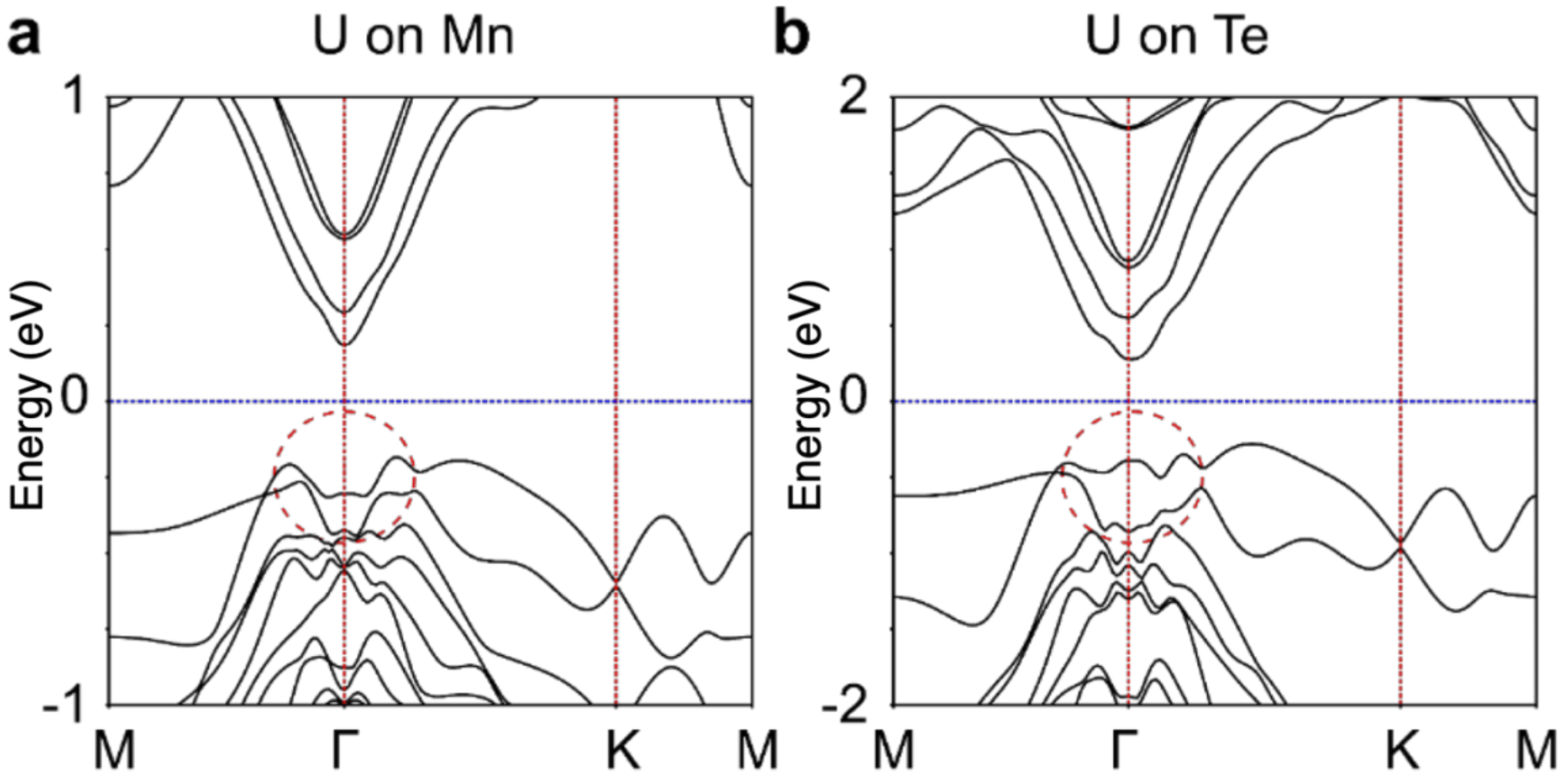

**Figure S4 |** Comparison between the GGA-PBE band structure of 1SL $MnBi_2Te_4$ with the effective on-site Hubbard correction U applied for **a**, Mn and **b**, Te atoms. The red dashed circles highlight the difference in band dispersion near the Fermi level.

**(b) Comparison between GGA-PBE and HSE**

Figure S5 plots the band structure of $MnBi_2Te_4$ 1SL, 2SL and 3SL obtained from DFT calculations using PBE-GGA (Fig. S5a-c) and those with using the hybrid HSE functional (Fig. S5d-f). No significant change between the two calculations is observed except for the magnitude of the band gap and an extra hump-like band ~ $E_B$ = 0.25-0.5 eV at the Γ point in GGA calculation, in the region circled in Fig. S5c-f.

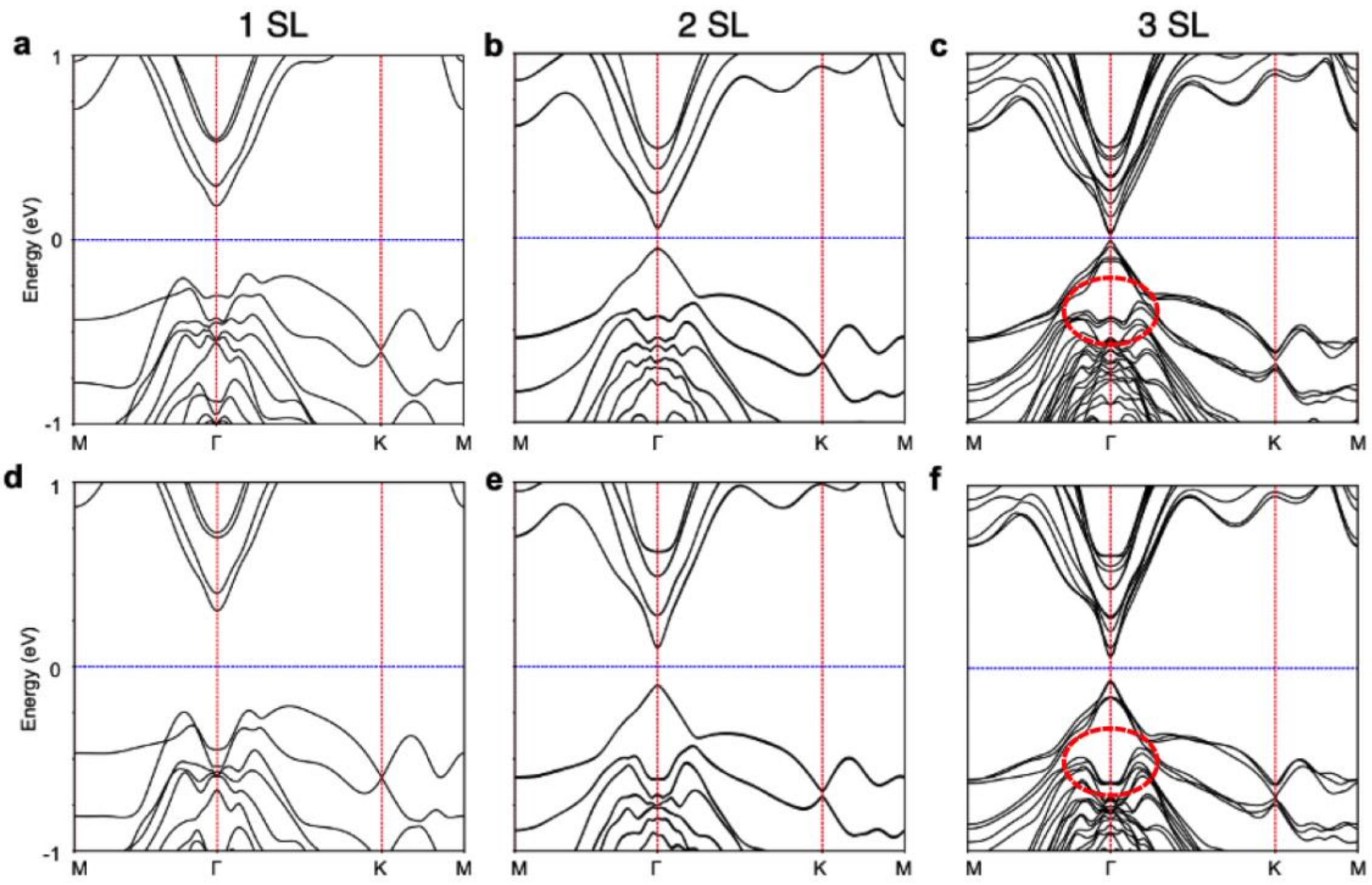

**Figure S5 |** Comparison between calculated **a-c**, GGA-PBE and **d-f**, HSE band structures for 1 SL, 2SL, and 3 SL.

## (c) Orbital character analysis

To have a better understanding about the orbital contribution to the bandgap and expected band inversion we plot the orbitals-resolved band structure for 1, 2, 3 and 5 SL $MnBi_2Te_4$ in Fig. S6a and b. Detailed analysis reveals that the orbital composition near the Fermi level is mainly composed of Bi 6*pz* and Te 5*pz*. A clear band inversion is observed for 3 SL and 5 SL consistent with non-trivial band topology and with previous theoretical findings.[9,10]

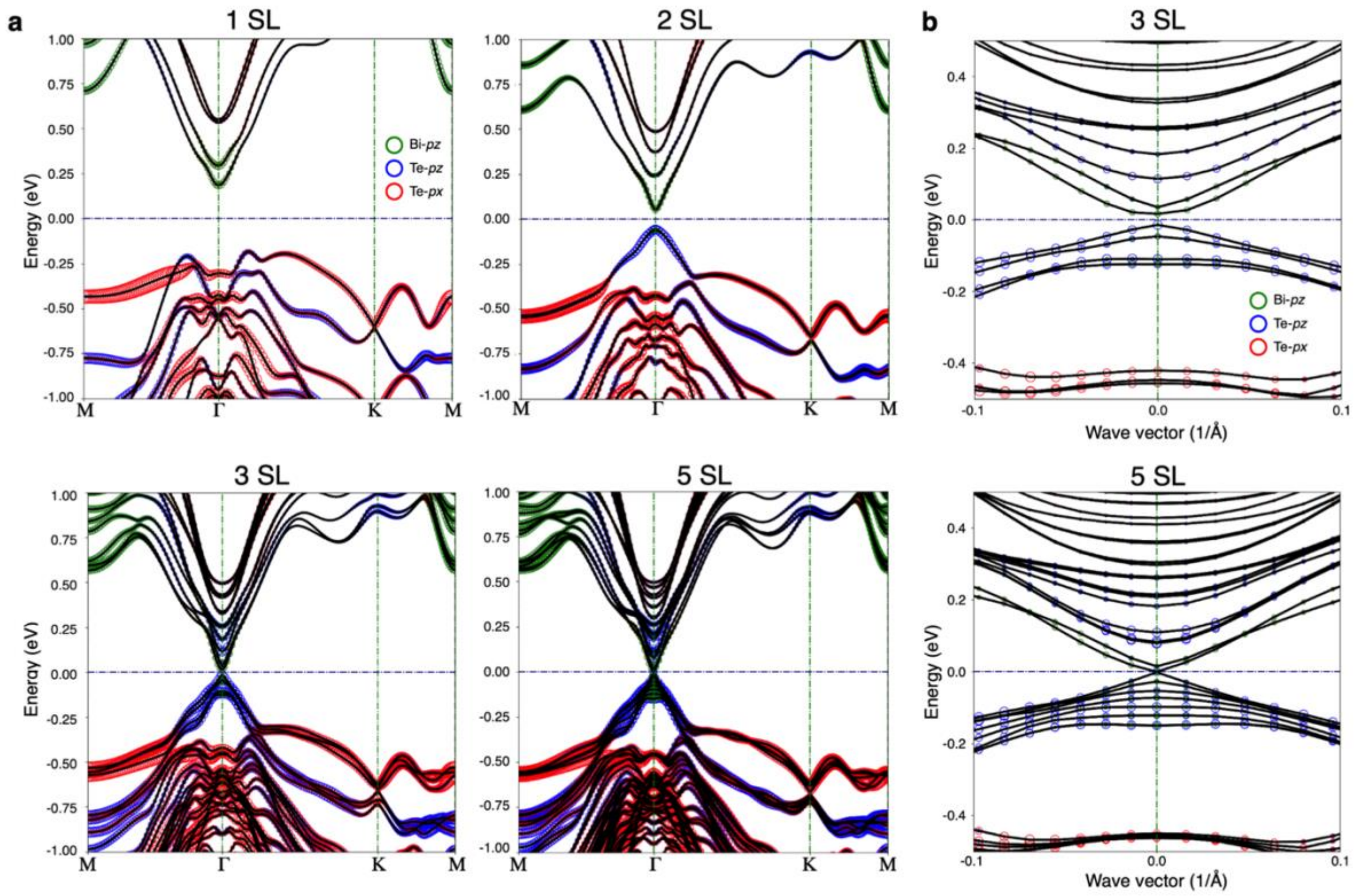

**Figure S6 | Orbital character calculation of 1 SL, 2 SL (top), 3 SL, and 5 SL (bottom). a**, Orbital character calculation showing the contributions of Bi *pz*, Te *pz* and Te *px* is shown for 1, 2, 3 and 5 SL $MnBi_2Te_4$ and **b**, Near-$E_F$ zoom in of 3 (top) and 5 SL (bottom).

## 5. Effective mass in 2 SL

The evolution of the 2 SL constant energy surfaces at different binding energies is shown in Fig. S7a. At the Fermi level, one can see the hexagonal shaped contour, similar to the typical warping effect in $Bi_2Te_3$. Moving down towards the conduction band minimum the warping becomes less pronounced and is more circular/isotropic. In order to calculate the electron effective mass ($m_e^*$) we fit the conduction band along the two high symmetry directions to a nearly free electron model given by $E = \frac{\hbar^2 k^2}{2m_e^*}$, to within a 0.15 Å$^{-1}$ region extending from the $\bar{\Gamma}$ point. Figure S7d-e re-plots the experimental data from (S7b and S7c) as $E$ vs. $k^2$, for the $\bar{\Gamma}\bar{M}$ and $\bar{\Gamma}\bar{K}$ directions respectively. This clearly demonstrates a linear relationship, i.e. a parabolic dispersion and constant effective mass, extending to momenta well away from the high symmetry point. Considering the uncertainty of each peak position due to the energy/momentum broadening, we obtain experimental $m_e^*$ values of $0.25 \pm 0.01 m_0$ for both $\bar{\Gamma}\bar{M}$ and $\bar{\Gamma}\bar{K}$.

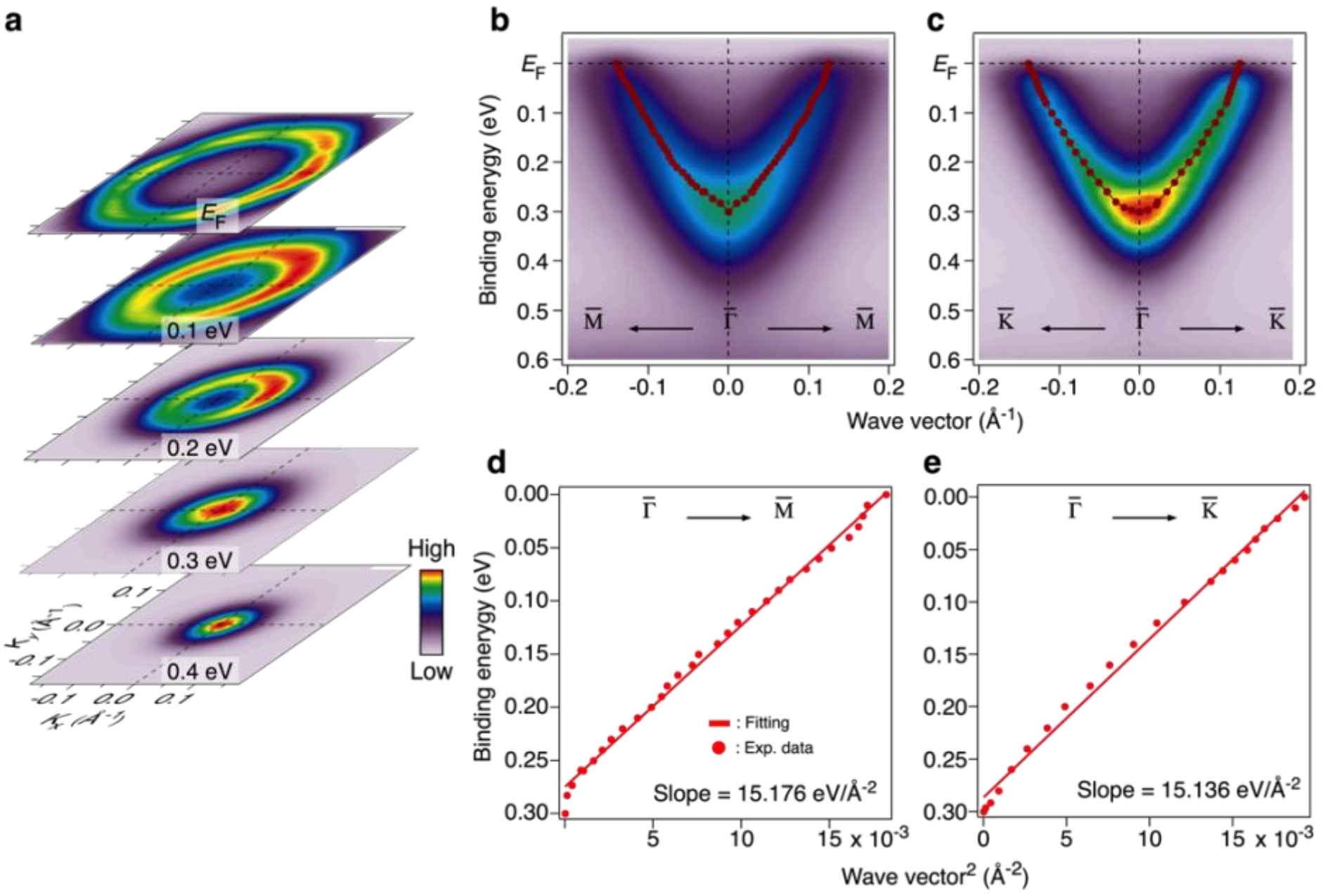

**Figure S7 | Near-$E_F$ ARPES intensity of 2 SL** at $h\nu$ = 50 eV and $T$ = 12 K**. a,** Constant energy contour stacked plots at different binding energies from Fermi level to 0.4eV with 0.1 eV step. Near-$E_F$ ARPES intensity of **b**, $\bar{\Gamma}\bar{M}$ and **c**, $\bar{\Gamma}\bar{K}$direction. Peak positions estimated from the EDCs and MDCs are marked as red dots and overlaid on the ARPES intensities. **d-e**, Plots of

binding energy as a function of $k^2$ showing a linear relationship with a slope of about 15.1 eV/Å$^{-2}$, yielding an effective mass of 0.25$m_0$.

## 6. Photon energy dependence of 5 SL

Near-$E_F$ ARPES dispersion of 5 SL has been observed with photon energy from 40 eV to 70 eV. As seen in Fig. S8, across a wide range of photon energies (40 - 70 eV), the X-shaped Dirac cone dispersion remains unchanged, but the intensity near the Dirac point dramatically changes around 50 eV to 60 eV due to a dominant spectral weight[11]. In this regard, to highlight the gap, we chose photon energy $h\nu$ = 50 eV where the spectral weight is optimized to see the lower Dirac bands.

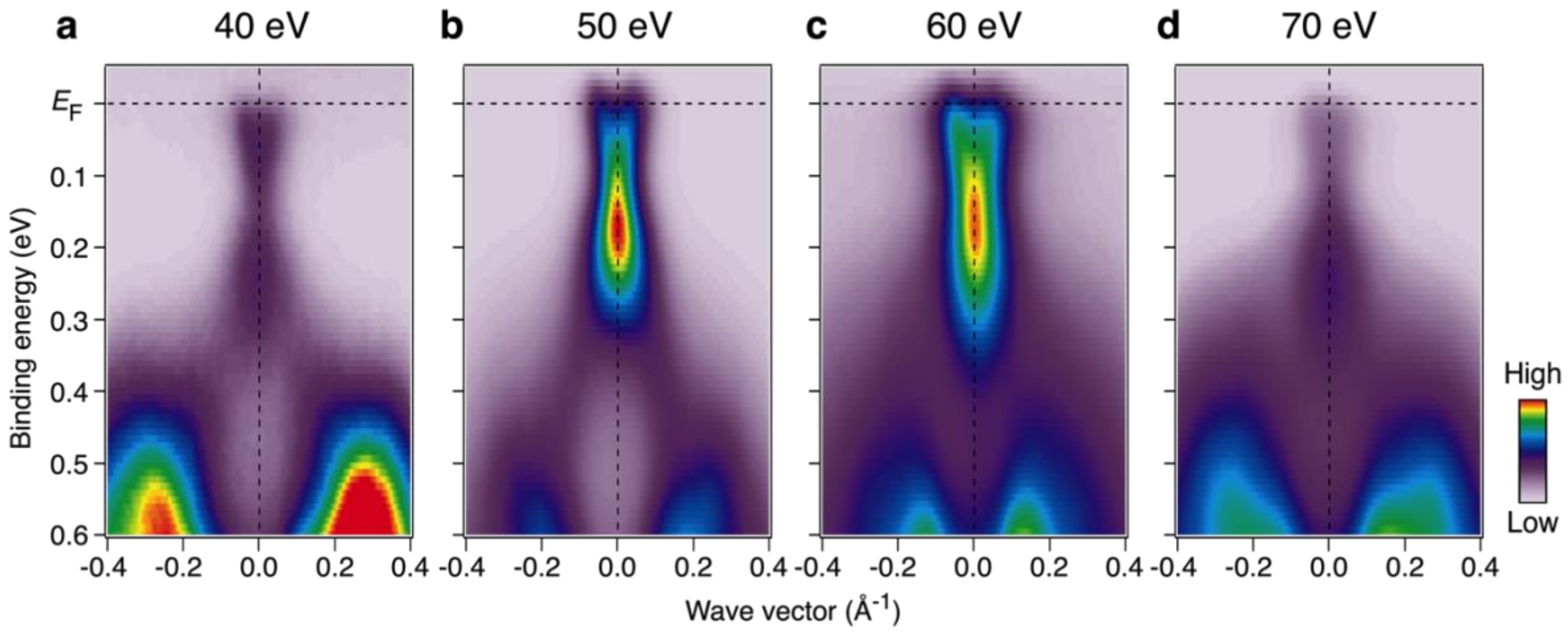

**Figure S8 |** Photon energy dependence of 5 SL at **a**, 40 eV, **b**, 50 eV **c**, 60 eV **d**, 70 eV taken at 13 K.

## 7. Temperature dependence of 5 SL

Near-$E_F$ ARPES intensity from 8 K to 33 K is depicted in Fig. S9a-b. When moving across $T_N$ from 8 K to 33 K, the bands expand in $k$ both in electron band and more significantly in the hole band below $E_B$~ 0.34 eV.

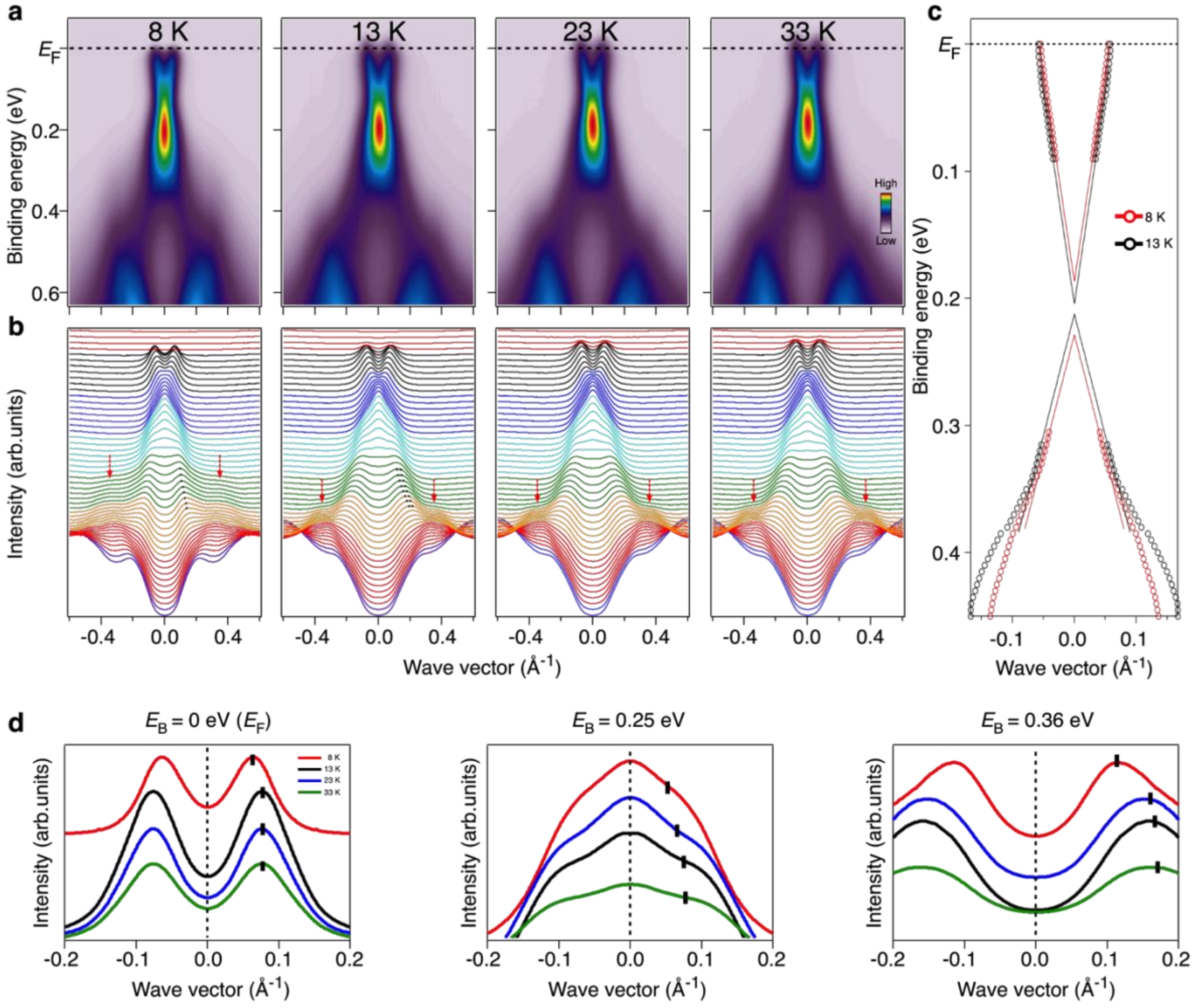

**Figure S9 | a**, near-$E_F$ ARPES intensity and **b**, MDCs of **a**, of 5 SL from 8 K to 33 K. MDCs with different colours designated as different binding energies. Black: $E_F$-0.09 eV, blue: 0.1-0.19 eV, sky blue: 0.2-0.29 eV, green: 0.3-0.39, orange: 0.4-0.49 eV, red: 0.5-0.59 eV, purple: 0.6-0.62 eV. The red and black arrows indicate valence band maxima and MDC peak maxima for the hole Dirac band, respectively. **c**, extracted MDC peak maxima showing the Fermi velocity of the electron and hole bands remains largely unchanged with increasing T, and linear extrapolated fits for 8 K (red) and 13 K (black) show the gap diminishes below 10 meV. **d**,

comparisons of extracted MDC maxima at variety temperatures, taken at binding energy $E_B$ = 0, 0.25, and 0.36 eV (left to right).

## 8. Photon-energy dependent gap of 5 SL

Figure S10 shows ARPES data measured with $h\nu$ = 40 eV at 13 K. In order to determine the bandgap, we first estimated the maxima of MDC peaks, shown as red dots in Fig. S10, then fitted the extracted MDC points with linear dispersion (solid white lines in Figure S10) indicative of Dirac bands. The linear bands result in a bandgap <10 meV, comparable to the bandgap of the data taken with $h\nu$ = 50 eV at 13 K in Fig. 5. In fact, the MDC points from data taken at 50 eV data in Fig. S10 (black circles) match the 40 eV data (red points) when overlaid. To confirm the reliability of MDC fitting, we also examined the EDC spectra in the $k$ range -0.02 to + 0.02 Å$^{-1}$. The EDC at $k$ = 0 Å$^{-1}$ (the red curve in the right half of Fig. S10) shows a clear V-shape in the Dirac point region and a distinct minimum, consistent with a gapless Dirac cone.

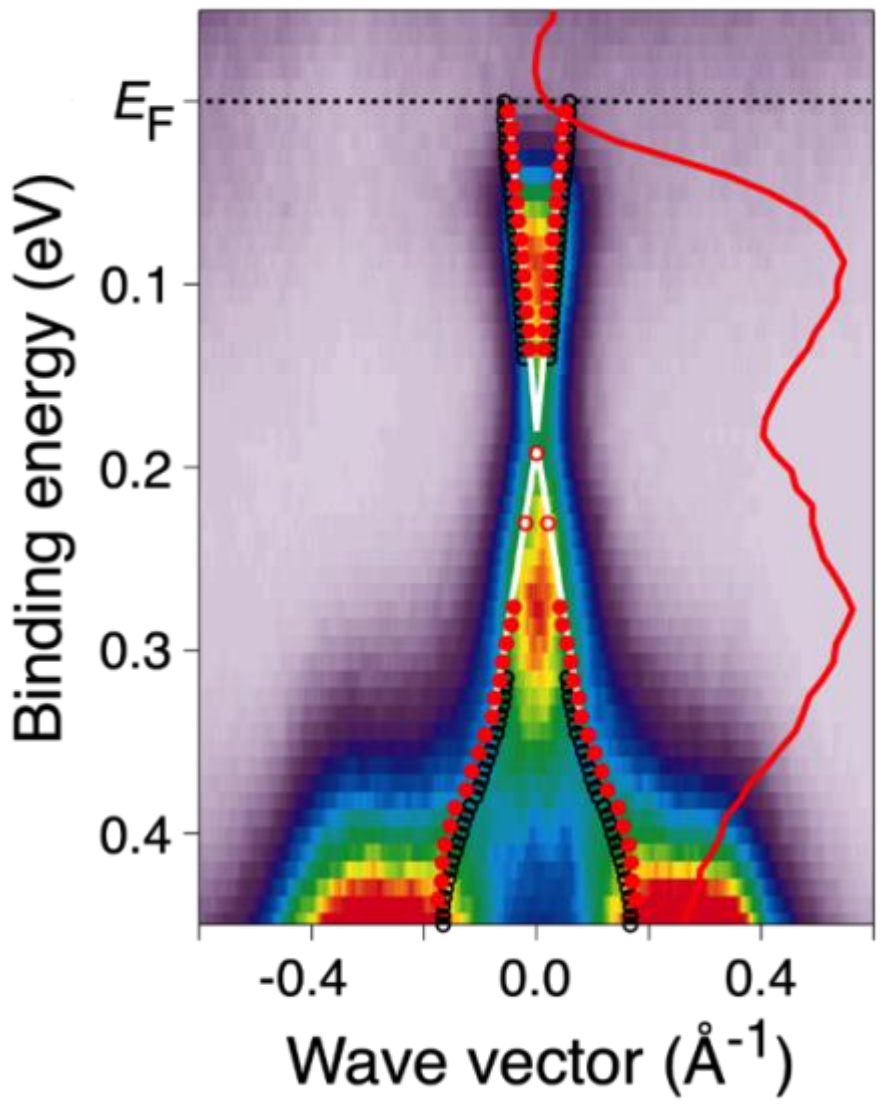

**Figure S10 |** Near-$E_F$ ARPES intensity of 5 SL measured with 40 eV at 13 K with overlaid fitting points from 40 eV (red) and 50 eV (black). The red solid dots and the red open dots are

MDC peak maxima and EDC peak maxima, respectively. The white solid lines are linear fittings represented for Dirac bands.